\documentclass[prd,aps,amsfonts,amssymb,mathrsfs,twocolumn,eqsecnum,amsmath,floatfix,showpacs,nofootinbib]{revtex4-1}
\usepackage{graphicx}
\usepackage{color}
\usepackage{dcolumn}
\renewcommand\a{\alpha}
\renewcommand\b{\beta}
\renewcommand\d{\delta}
\renewcommand\k{\kappa}
\renewcommand\l{\lambda}
\renewcommand\r{\rho}

\renewcommand\t{\tau}

\newcommand\g{\gamma}
\newcommand\z{\zeta}
\newcommand\m{\mu}
\newcommand\n{\nu}

\newcommand\p{\pi}

\newcommand\s{\sigma}

\newcommand\w{\eta}

\newcommand\ve{\varepsilon}

\renewcommand\L{\Lambda}

\renewcommand\H{\Theta}
\newcommand\D{\Delta}

\newcommand{\fig}[1]{Fig.~\ref{#1}}
\newcommand{\eq}[1]{Eq.~(\ref{#1})}

\newcommand\lb{\left(}
\newcommand\rb{\right)}
\newcommand\ls{\left[}
\newcommand\rs{\right]}

\newcommand{\lan}{\langle}
\newcommand{\ran}{\rangle}
\newcommand\ua{\uparrow}
\newcommand\da{\downarrow}

\newcommand{\non}{\nonumber\\}
\newcommand\pt{\partial}

\newcommand{\bb}{{\mathbf b}}

\newcommand{\bp}{{\mathbf p}}
\newcommand{\bk}{{\mathbf k}}
\newcommand{\bq}{{\mathbf q}}
\newcommand{\bxh}{{\hat{\mathbf x}}}
\newcommand{\byh}{{\hat{\mathbf y}}}
\newcommand{\bzh}{{\hat{\mathbf z}}}
\newcommand{\bxt}{{{\mathbf x}_T}}

\newcommand{\idp}[2]{\int\frac{d^{\,#1}#2}{(2\p)^#1}}

\newcommand{\etal}{\emph{et al.}}

\newcommand{\gton}{\mathrel{\lower.9ex \hbox{$\stackrel{\displaystyle
>}{\sim}$}}}
\newcommand{\lton}{\mathrel{\lower.9ex \hbox{$\stackrel{\displaystyle
<}{\sim}$}}}

\begin{document}

\title{Quark Polarization in a Viscous Quark-Gluon Plasma}
\author{Xu-Guang Huang$^{1,2}$}\email{xhuang@th.physik.uni-frankfurt.de}
\author{Pasi Huovinen$^2$}\email{huovinen@th.physik.uni-frankfurt.de}
\author{Xin-Nian Wang$^{3,4,2}$}\email{XNWang@lbl.gov}
\affiliation{$^1$ Frankfurt Institute for Advanced Studies, D-60438 Frankfurt am Main, Germany\\
$^2$Institut f\"ur Theoretische Physik, Goethe-Universit\"at, D-60438 Frankfurt am Main, Germany\\
$^3$Institute of Particle Physics, Central China Normal University, Wuhan, 430079,China\\
$^4$Nuclear Science Division, MS 70R0319, Lawrence Berkeley National Laboratory,
    Berkeley, California 94720, USA}

\date{\today}

\begin{abstract}
Quarks produced in the early stage of noncentral heavy-ion
collisions could develop a global spin polarization along the
opposite direction of the reaction plane due to the spin-orbital
coupling via interaction in a medium that has finite longitudinal flow shear along the direction
of the impact parameter. We study how such polarization evolves via multiple scattering
in a viscous quark-gluon plasma with an initial laminar flow.
The final polarization is found to be sensitive to the viscosity and the initial shear of local longitudinal flow.
\end{abstract} 

\maketitle

\section {Introduction}\label{1}
The observed jet quenching and collective phenomena in high-energy
heavy-ion collisions at the Relativistic Heavy Ion Collider (RHIC)
provide strong evidence of the formation of strongly coupled
quark-gluon plasma (QGP)~\cite{nucl-th/0405013,hep-ph/0405125}: The
strong quenching of high transverse momentum jets is understood
to be caused by parton energy loss induced by multiple collisions of
the leading parton with color charges in the thermal
medium~\cite{nucl-th/9306003,hep-ph/9607440,hep-ph/0005129,nucl-th/0006010,hep-ph/9607355,hep-ph/0005044};
the observed collective flow in the final bulk hadron spectra indicates
a hydrodynamic behavior of the initial dense matter as an almost perfect fluid with a very small shear
viscosity~\cite{arXiv:0804.4015,arXiv:1011.2783}, $\w/s\lesssim
0.5$. The large jet transport parameter from the observed strong jet quenching and small shear viscosity
inferred from the collective flow can be connected to each other through a transport process  in
a strongly coupled system~\cite{Majumder:2007zh}. They both describe the ability of the medium
partons to transfer momentum via strong interaction in QCD and maintain local equilibrium. Globally,
such transport processes help to dissipate variations of flow velocities and thus will reduce the
anisotropic flow, which is driven by the initial geometric anisotropy~\cite{arXiv:0804.4015,arXiv:1011.2783}.
In this paper, we discuss the possibility of global quark spin polarization caused by such
transport processes in noncentral high-energy heavy-ion collisions.

It was first proposed by Liang and Wang~\cite{nucl-th/0410079} that
global quark polarization could occur in the QGP formed in a
noncentral heavy-ion collision. They argued that at a finite
impact parameter, the initial partons produced in the collision can develop
a longitudinal fluid shear distribution representing local relative
orbital angular momentum (OAM) in the same direction as the global OAM
of the noncentral nucleus-nucleus collisions. Since interaction via one-gluon exchange in
QCD contains a spin-orbital coupling, the OAM could cause a global spin-polarization of quarks
and antiquarks in the direction parallel to the OAM. Such a global
(anti)quark polarization should have many observable consequences
such as global hyperon polarization~\cite{nucl-th/0410079,arXiv:0708.0035},
vector meson spin alignment~\cite{nucl-th/0410079,nucl-th/0411101}, and the
emission of circularly polarized photons~\cite{arXiv:0710.5700}.
Predictions have been
made~\cite{nucl-th/0410079,nucl-th/0411101,arXiv:0705.2852,arXiv:0710.2943,arXiv:0710.5700}
for these measurable quantities as functions of the global quark
polarization $P_q$. Experimental measurements of the
$\L$ hyperon polarization with respect to the reaction plane at
RHIC~\cite{nucl-ex/0510069,nucl-ex/0605035,nucl-ex/0608034,nucl-ex/0701034,arXiv:0705.1691,arXiv:0801.1729,Chen:2007zzq,Chen:2008zzg}
place a limit $|P_{\L,\bar{\L}}|\lesssim 0.02$~\cite{nucl-ex/0605035,Chen:2007zzq}.
 Such a limit puts a stringent test on both the initial shear of longitudinal flow in noncentral heavy-ion collisions~\cite{arXiv:0710.2943}
 as well as the time evolution of the polarization through transport processes.

The estimates of the global quark polarization in
Ref.~\cite{nucl-th/0410079} and in subsequent
studies~\cite{arXiv:0705.2852,arXiv:0710.2943,arXiv:0801.2296,gao2007} were
all obtained by considering the polarization process for a single
scattering between quarks and thermal partons. However, one should
consider the effect of the multiple scattering and expect that the
quarks will be progressively polarized through multiple scattering.
Furthermore, with the minimum values of shear viscosity $\eta/s\geq 1/4\pi$ in QGP imposed by
the quantum limit, the local momentum shear, $dp_z/dx$, of the fluid, that is, the local
OAM of interacting parton pairs, will decay with time.
This will lead to a nontrivial time evolution of quark polarization $P$ depending
on the shear viscosity of the QGP matter
and the final state observed global polarization could serve as a
viscometer of QGP. In this paper, we focus on these two issues
with a simple and yet interesting hydrodynamic evolution of a
relativistic laminar flow between two frictionless impenetrable
walls.

The rest of the paper is organized as follows. In Sec.~\ref{2}, we
extend the calculation in Ref.~\cite{nucl-th/0410079} to the case of
scattering of an initially polarized quark in a static potential model. In
Sec.~\ref{3}, we study the relativistic laminar
flow and compute the decay of the longitudinal momentum gradient.
The results of Sec.~\ref{2} and Sec.~\ref{3} are applied to
Sec.~\ref{4} to study the time evolution of the quark polarization.

\section {Polarization of Initially Polarized quarks}\label{2}
\begin{figure}[!htb]
\begin{center}
\includegraphics[width=7.5cm]{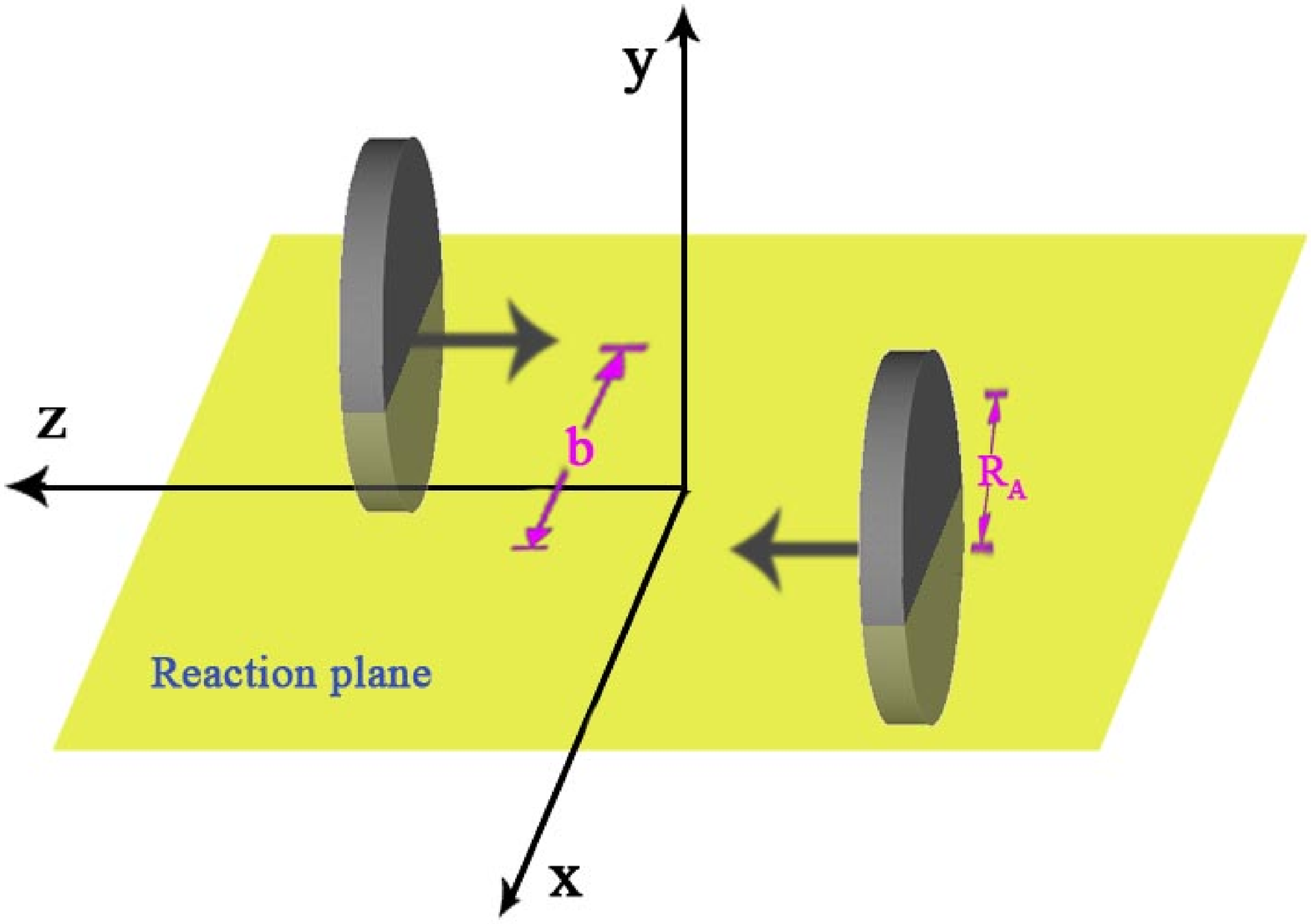}
\includegraphics[width=7.5cm]{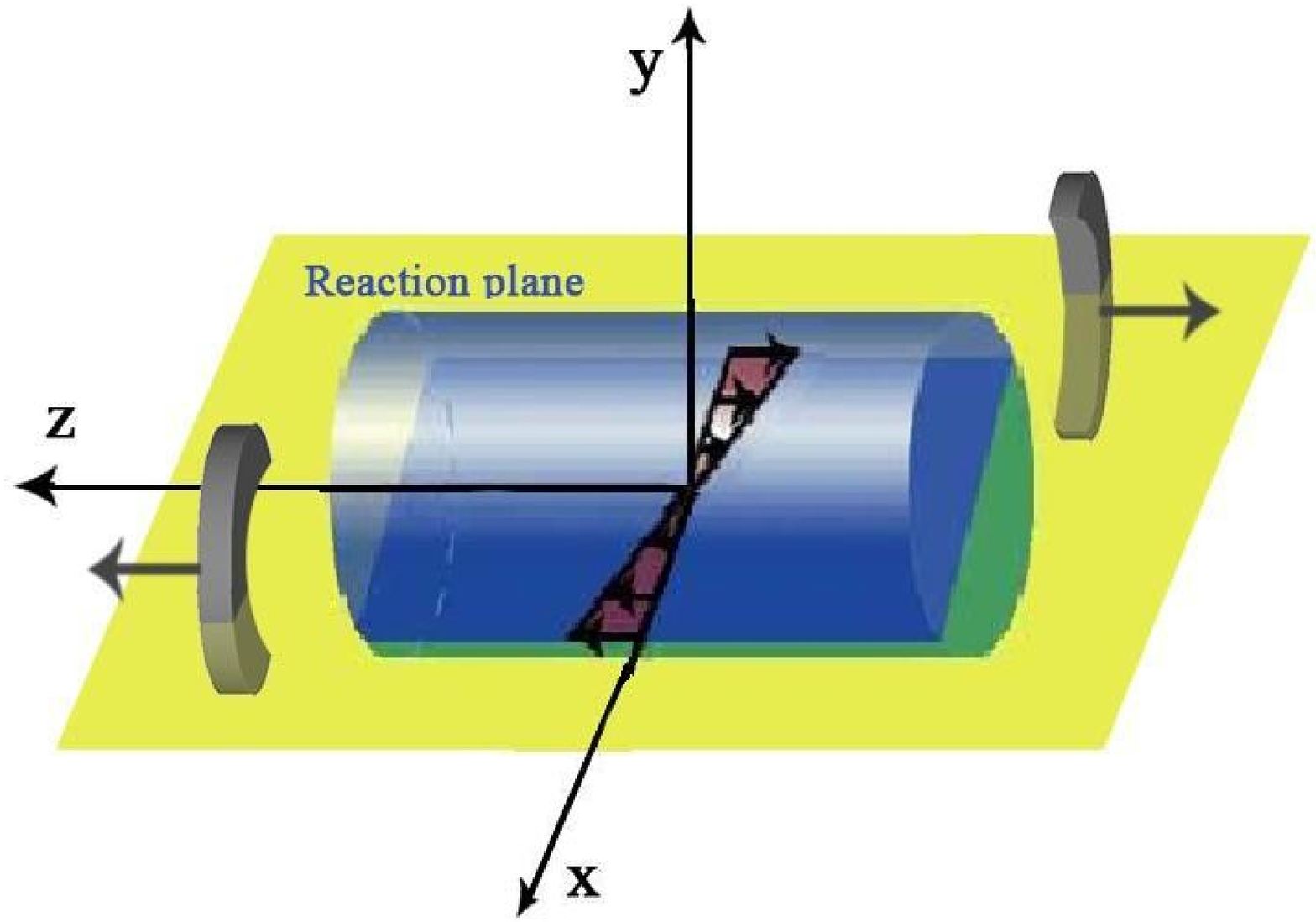}
\caption{(Color online) Illustration of noncentral collisions with
impact parameter $\bb$ of two heavy nuclei with radii $R_A$. The
global angular momentum of the produced matter is along $-\byh$,
opposite to the reaction plane.} \label{rhic}
\end{center}
\end{figure}

We consider two colliding nuclei with the beam projectile moving
in the direction of $\bzh$ and the impact parameter $\bb$ defined as
the transverse distance of the projectile from the target nucleus
along the $\bxh$ direction as illustrated in the upper panel of
\fig{rhic}. The direction $\byh$ defines the reaction plane,
$\byh\propto\bzh\times\bxh$. The initial OAM of these two colliding
nuclei is along the direction opposite to the reaction plane and
could be very large. Given $1\, {\rm fm}<b<10\, {\rm fm}$, the
initial OAM $L_0\simeq Ab\sqrt{s}/2$ is roughly $10^5\lesssim
L_0\lesssim10^6$ for Au-Au collisions at RHIC energy $\sqrt{s}=200$
GeV and $3\times10^6\lesssim L_0\lesssim3\times10^7$ for Pb-Pb
collision at Large Hadron Collider (LHC) energy $\sqrt{s}=5.5$ TeV. Because of the unequal
local number density of the participant projectile and target
nucleons at various transverse positions, some fraction of
this large OAM could be transferred into the produced QGP matter in the
overlapping region. Such global angular momentum, however, would never lead
to a collective rotation of the system since there is no strong binding or
attractive interaction in the partonic interaction at high energy. Instead, it
could be manifested in the finite transverse (along $\bxh$) gradient
of the average longitudinal momentum $p_z$ per produced parton due
to the partonic interaction at high energy (see the lower panel of
\fig{rhic}). Given the range of interaction $\D x$, two colliding
partons will have relative longitudinal momentum $\D p_z=\D
xdp_z/dx$ with relative OAM $L_y\sim-\D x \D p_z$. This relative OAM
will lead to global quark polarization along $-\byh$ through the
spin-orbital coupling in QCD. This is essentially the argument that
was first proposed in Ref.~\cite{nucl-th/0410079}. It was found
that the quark polarization via a single scattering with given
relative momentum $p$ reads
\begin{eqnarray}
\label{polar1}
P\equiv\frac{\D\s}{\s}\equiv\frac{\s_\ua-\s_\da}{\s_\ua+\s_\da}=-\frac{\p
\m p}{2E(E+m)},
\end{eqnarray}
where $\s_s,s=\ua,\da$ is the cross section of final quark with spin
$s$ along $\byh$, $m$ is the mass of interacting quark, and $\m$ is
the Debye screening mass of longitudinal gluon,
$\m^2=g^2(N_c+N_f/2)T^2/3$. The initial relative momentum $p$ can be
estimated as $p\simeq\D x dp_z/dx$ with $\D x\sim\m^{-1}$ being the
characteristic range of interaction. Then $p/\m$ is nothing but the
relative orbital angular momentum between the scattering quarks,
$L_y\sim - p/\m$. In the non-relativistic limit for massive quarks,
 $P$ is proportional to the spin-orbital coupling energy
  $P \propto E_{LS}/\mu$ where $E_{LS}=(\vec L\cdot \vec S)(dV_{0}/dr)/rm^{2}$ and $(dV_{0}/dr)/r\sim \mu^{3}$
  with typical interaction range $r\sim 1/\mu$.

The estimates in
Refs.~\cite{nucl-th/0410079,arXiv:0705.2852}, and~\cite{arXiv:0710.2943} were
based on the assumption that the initial quarks are not polarized.
In order to discuss the time evolution of the quark polarization via multiple scattering, one must calculate
the quark-quark cross section of initially polarized quarks. Let the
fraction of initial quarks of spin $\l_i/2$ along $\byh$ be
$R_{\l_i}=(1+\l_iP_i)/2$ with $P_i$ being the initial polarization.
The identity $R_++R_-=1$ must hold. Consider a quark with initial
relative four-momentum $p^\m=(E,\bp)$ and spin $\l_i/2$ scattering with
a virtual gluon and resulting in final spin $\l_f/2$; the cross
section with fixed impact parameter $\bxt$ is
\begin{eqnarray}
\label{sig}
\frac{d\s_{\l_f}}{d^2\bxt}&=&C_T\sum_{\l_i}\idp{2}{\bq_T}\idp{2}{\bk_T}e^{i(\bk_T-\bq_T)\cdot\bxt}\non&&\times\,
R_{\l_i}\,{\cal I}_{\l_f\l_i}(\bk_T,\bq_T,E),\non{\cal
I}_{\l_f\l_i}&\equiv&{\cal M}_{\l_f\l_i}(\bq_T,E){\cal
M}^*_{\l_f\l_i}(\bk_T,E),\non {\cal
M}_{\l_f\l_i}(\bq_T,E)&=&\frac{g}{2E}{\bar
u}_{\l_f}(p_q)A\!\!\!/(\bq_T)u_{\l_i}(p),
\end{eqnarray}
where $C_T=2/9$ is the color factor associated with the target,
$\bq_T$ ($\bk_T$) is the transverse momentum transfer from the
virtual gluon to quark, and $p_{q(k)}^\m$ is the final four-momentum of
quark, $p_{q(k)}^\m=p^\m+[0,\bq_T(\bk_T)]$. We use the screened
static potential model to calculate ${\cal
  M}_{\l_f\l_i}$ in which $A^\m=(A^0,{\bf 0})$ with
$A^0(q_T)=g/(q_T^2+\m^2)$~\cite{nucl-th/9306003}.

For small angle scattering (which is justified when the relative
longitudinal momentum $\bf p$ is large), $q_T,k_T\sim\m\ll E$,
one finds
\begin{eqnarray}
\label{I} {\cal
I}_{\l_f\l_i}&\approx&\frac{g^2}{2}A_0(q_T)A_0(k_T)\Big\{1+\l_i\l_f\non&&+\frac{1}{2E(E+m)}\big[
(1+\l_i\l_f)\bp\cdot(\bq_T+\bk_T)\non&&+i(\l_i+\l_f)\bp\cdot\byh\times(\bk_T-\bq_T)\big]
\Big\}.
\end{eqnarray}


From \eq{sig} and \eq{I}, it is evident that the polarization will
not change if one averages the cross section over all the possible
directions of the parton impact parameter $\bxt$. However, in
noncentral heavy-ion collisions, the local relative OAM $L_y$ provides a preferred average reaction plane for
parton collisions. This will lead to a global quark polarization
opposite to the reaction plane of nucleus-nucleus collisions. This
conclusion should not depend on our perturbative treatment of parton
scattering as far as the effective interaction is mediated by the
vector coupling in QCD. Therefore, we average over the upper
half-$xy$-plane with $x>0$, that is, average over the relative angle
between parton $\bxt$ and the nuclear impact parameter $\bb$ from
$-\p/2$ to $\p/2$ and over $x_T$. To do this, we use the identity
\begin{eqnarray}
\int_{x>0}d^2\bxt e^{i(\bk_T-\bq_T)\cdot\bxt}=\frac{2\p
i\d(k_y-q_y)}{k_x-q_x+i0^+}.
\end{eqnarray}
Then the total unpolarized cross section reads,
\begin{eqnarray}
\label{crosec}
\s&\equiv&\int_{x>0}d^2\bxt\frac{d\s}{d^2\bxt} \equiv \int_{x>0}d^2\bxt\ls\frac{d\s_+}{d^2\bxt}+\frac{d\s_-}{d^2\bxt}\rs \non
&=&\int_0^\infty
dq_T q_T\frac{C_Tg^4}{4\p(q_T^2+\m^2)^2}\non&&\times\ls1-P_i\frac{
p\sqrt{q_T^2+\m^2}K(q_T/\sqrt{q_T^2+\m^2})}{\p
E(E+m)}\rs\non
&=&\frac{C_Tg^4}{8\p\m^2}\ls1-P_i\frac{\p\m
p}{2E(E+m)}\rs,
\end{eqnarray}
and the polarized cross section reads
\begin{eqnarray}
\label{dcrosec}
\D\s&\equiv&\int_{x>0}d^2\bxt\frac{d\D\s}{d^2\bxt} \equiv \int_{x>0}d^2\bxt\ls
\frac{d\s_+}{d^2\bxt}-\frac{d\s_-}{d^2\bxt}\rs\non
&=&\int_0^\infty
dq_T q_T\frac{C_Tg^4}{4\p(q_T^2+\m^2)^2}\non&&\times\ls P_i-\frac{
p\sqrt{q_T^2+\m^2}K(q_T/\sqrt{q_T^2+\m^2})}{\p
E(E+m)}\rs\non&=&\frac{C_Tg^4}{8\p\m^2}\ls P_i-\frac{\p\m
p}{2E(E+m)}\rs,
\end{eqnarray}
where $K(x)$ is the complete elliptic integral of the first kind.
The final global quark polarization is then
\begin{eqnarray}
\label{pf} P_f
&=&P_i-\frac{(1-P_i^2)\p\m p}{2E(E+m)-P_i\p\m p}.
\label{pol}
\end{eqnarray}

It is also useful to get the transverse momentum dependence of the
quark polarization. From \eq{crosec} and \eq{dcrosec}, we read out
the differential cross sections,
\begin{eqnarray}
\frac{d\D\s}{dq_T}&=&q_T\frac{C_Tg^4}{4\p(q_T^2+\m^2)^2}\non&&\times\ls
P_i-\frac{ p\sqrt{q_T^2+\m^2}K(q_T/\sqrt{q_T^2+\m^2})}{\p
E(E+m)}\rs,\\
\frac{d\s}{dq_T}&=&q_T\frac{C_Tg^4}{4\p(q_T^2+\m^2)^2}\non&&\times\ls
1-\frac{P_i p\sqrt{q_T^2+\m^2}K(q_T/\sqrt{q_T^2+\m^2})}{\p
E(E+m)}\rs. \non
\end{eqnarray}
The transverse-momentum-dependent polarization (TMDP) defined as
$P_f(q_T)\equiv (d\D\s/dq_T)/(d\s/dq_T)$ now reads,
\begin{eqnarray}
\label{diff_pol} P_f(q_T)=\frac{\p
E(E+m)P_i-p\sqrt{q_T^2+\m^2}K(q_T/\sqrt{q_T^2+\m^2})}{\p
E(E+m)-P_ip\sqrt{q_T^2+\m^2}K(q_T/\sqrt{q_T^2+\m^2})}.\non
\end{eqnarray}

\begin{figure}[!htb]
\begin{center}
\includegraphics[width=7.5cm]{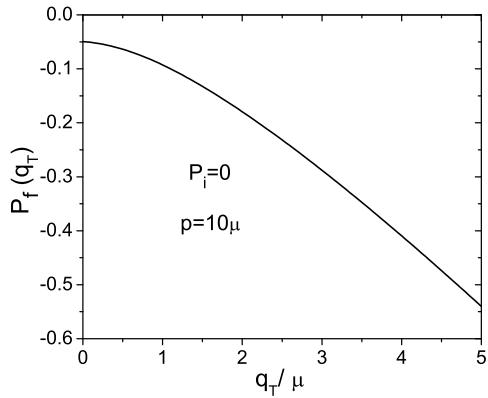}
\caption{The TMDP  as a function of transverse momentum in unit of
$\m$. The initial relative longitudinal momentum is chosen to be
$p=10\m$.} \label{dpfdqt}
\end{center}
\end{figure}

Some discussions are in order. (1) If the initial quark is unpolarized,
$P_i=0$, we recover the result of Ref.~\cite{nucl-th/0410079}. (2)
Because the denominator is always positive in the right-hand-side (RHS) of \eq{pf} for
high relative longitudinal momentum (i.e., when small angle
approximation is applicable), we always have $P_f\leqslant P_i$. Therefore
scattered quarks always prefer to be polarized along $-\byh$ direction.
(3) The scattering matrix elements ${\cal{I}}_{\l_f\l_i}$ with spin
flipping ($\l_f=-\l_i$) are zero according to \eq{I}, so there is no flipping of
quark's spin via the scattering under this small angle approximation.
The polarization in the final state is caused by the
larger cross section of quarks with spin up relative to quarks with
spin down. This will lead to the conclusion that if the initial
quark is completely polarized, $P_i=\pm1$, we must have $P_f=P_i$.
This is indeed the case expressed in \eq{pf} when $P_i=\pm1$. (4) The
quark polarization has a remarkable transverse momentum dependence,
as shown in \eq{diff_pol}. Figure \ref{dpfdqt} shows the typical behavior
of TMDP as a function of the transverse momentum with given
$p=10\m$. The polarization grows with the transverse momentum due to quark-quark
scattering. In principle, the $\L$-hyperon polarization should
have similar transverse momentum dependence, although as we
mentioned in Sec.~\ref{1} it is not trivial to construct a correspondence
between quark polarization and hadron polarization.

\section {Relativistic laminar flow}\label{3}
\begin{figure}[!htb]
\begin{center}
\includegraphics[width=8cm]{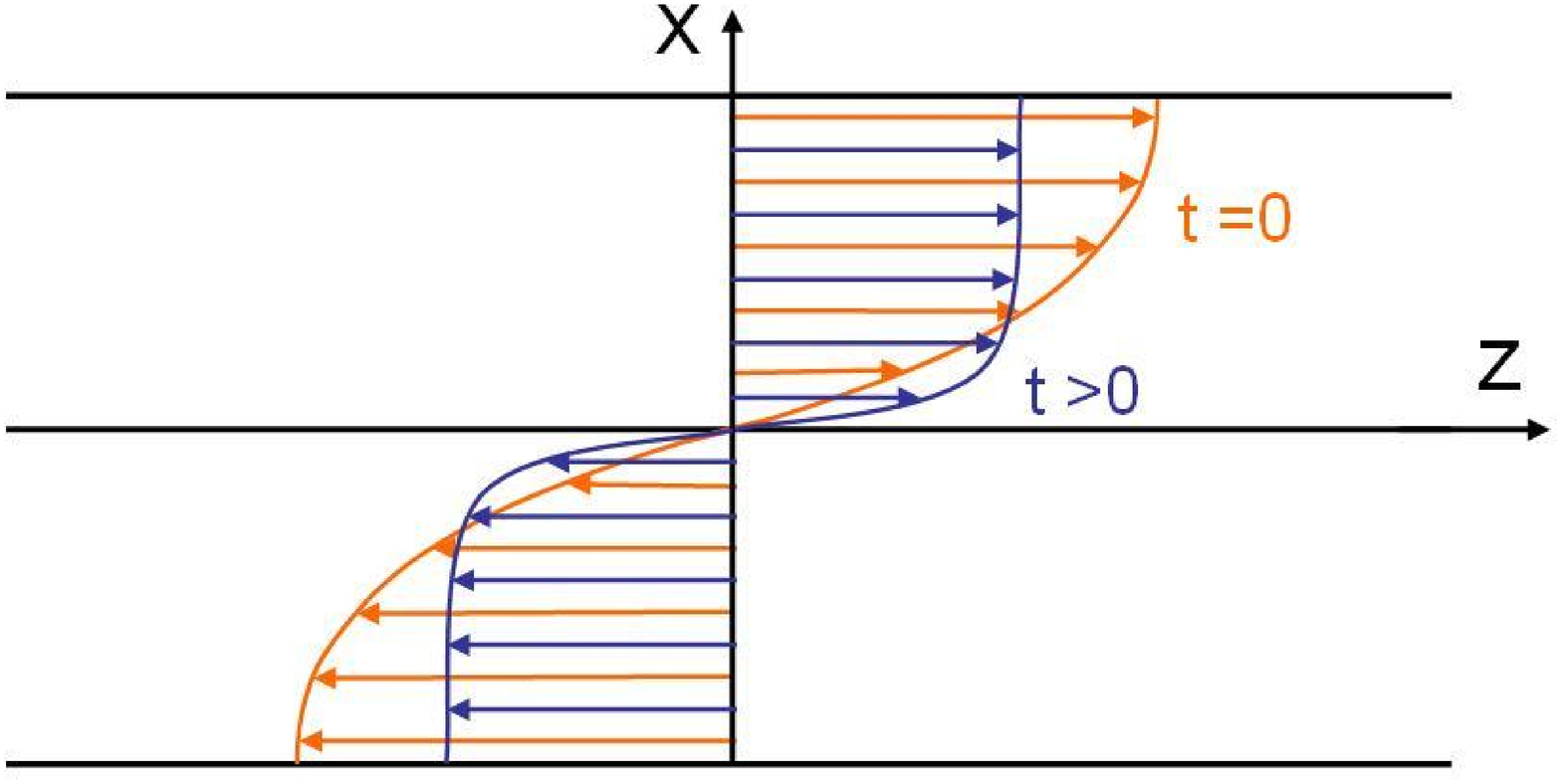}
\caption{(Color online) Illustration of the velocity profiles of the
relativistic laminar flow.} \label{laminar}
\end{center}
\end{figure}

Before discussing how the quark polarization evolves in a viscous
QGP due to multiple scattering, we have to know how the QGP itself evolves
through either transport model or viscous hydrodynamical
model~\cite{arXiv:0804.4015,arXiv:1011.2783,arXiv:0710.5932,arXiv:0709.0742,arXiv:0806.1367,arXiv:0903.3595,Schenke:2010rr},
Moreover, we also have to know the initial profile of the longitudinal flow field. In the discussion
in Sec.~\ref{2}, we simply followed Ref.~\cite{nucl-th/0410079}, and
assumed that nothing depends on the longitudinal position in the
system. In such a case, the finite angular momentum must lead to a
velocity profile depicted in Fig.~\ref{laminar} (see
Ref.~\cite{arXiv:0711.1253} for a discussion of possible
consequences of such a profile). On the other hand, another extreme
is to assume that $dv_z/dx\equiv 0$ everywhere, but the angular
momentum is carried by the matter distribution in the reaction
plane; see Ref.~\cite{nucl-th/0505004} for illustration.

To study the effect of viscosity on the decay of the
local angular momentum, we consider a simple laminar flow without driving force
between two frictionless (free-slip flow) impenetrable walls. We
assume the walls are infinitely large and separated by a distance
$2h$. To make dimensions relevant for a heavy-ion collision, we set
$h=5$ fm. Such a scenario might be far from the real longitudinal flow profile in high-energy
heavy-ion collisions, but it will be very illustrative for our  study here.
We further assume that the flow profile has no
longitudinal variation and the system has a reflection symmetry
respect to the $yz$-plane. We study two cases: One with no expansion, and another
 with boost-invariant expansion in $y$ direction, that is, with flow profile $v_y = y/t$.
In both cases, the flow four-velocity in the reaction plane has the general form $u^\m=(\g,\g v_x,
0,\g v_z)$ with $\g\equiv1/\sqrt{1-v_x^2-v_z^2}$ and $v_{x,z}(t,x)$
being the $x$ and $z$ components of the three-velocity.

As is well-known, the relativistic Navier-Stokes hydrodynamics is
unstable and provides a possibility for acausal signal
velocities~\cite{Hiscock:1983zz}. Therefore, we use the second-order
theory by Israel and Stewart~\cite{Israel:1979wp} instead. Although
hydrodynamics has been widely used to model the heavy-ion
collisions, as far as we know, there is no literature discussing
the relativistic laminar flow.

If there are no conserved charges, the hydrodynamical equations of
motion are given by the conservation of energy and momentum
\begin{eqnarray}
\pt_\m T^{\m\n}&=&0,
\end{eqnarray}
where $T^{\m\n}\equiv(\ve+\H)u^\m u^\n-\H g^{\m\n}+\p^{\m\n}$ is the
energy-momentum tensor, $\ve$ is the energy density, $\H$ is the
pressure\footnote{Note that
      since we used $P$ to denote polarization, to avoid confusion, we do not use it to
      denote pressure.}, and $\p^{\m\n}$ is the
shear stress tensor. To close the set of differential equations, one
also needs to specify an equation of state (EOS) $\ve=\ve(\H ) $.
For simplicity, we use the ideal gas EOS $\ve=3\H$.

In its simplest form, Israel-Stewart hydrodynamics means that
instead of being directly proportional to the velocity gradients,
the shear stress tensor is a dynamical variable,
which relaxes toward the Navier-Stokes value on its relaxation time
$\t_\p$:
\begin{eqnarray}
D\p^{\m\n}=-\frac{1}{\t_\p}\lb\p^{\m\n}-2\w\nabla^{\lan\m}u^{\n\ran}\rb-2\p^{\k(\m}u^{\n)}Du_\k,
\end{eqnarray}
where $D\equiv u^\l\pt_\l$, $A^{(\m\n)}\equiv(A^{\m\n}+A^{\n\m})/2$,
$A^{\lan\m\n\ran}\equiv[\D^{(\m}_\a\D^{\n)}_\b-\frac{1}{3}\D^{\m\n}\D_{\a\b}]
A^{\a\b}$, $\nabla^\m\equiv\pt^\m-u^\m u^\n\pt_\n$, $\D^{\m\n}\equiv
g^{\m\n}-u^\m u^\n$ and $\w$ is the shear viscosity coefficient. The
last term is required to keep the shear stress tensor orthogonal to
the flow velocity in all circumstances. This is the so-called
truncated Israel-Stewart equation. Although there are more terms in
a complete Israel-Stewart equation, for our purpose here, the
truncated one is sufficient.

The relaxation time is given by~\cite{Israel:1979wp}
\begin{eqnarray}
\t_\p=2\w\b_2,
\end{eqnarray}
which is dependent on the shear viscosity and another coefficient
$\b_2$. For massless Boltzmann particles, the kinetic theory
gives~\cite{Israel:1979wp}
\begin{eqnarray}
\b_2=\frac{3}{4\H}.
\end{eqnarray}
If there is no phase transition, it is expected that
$\b_2$ for Fermion and Boson gases have only minor modification from
$\b_{2}$ for Boltzmann gas at high
temperature~\cite{arXiv:0712.2451,natsuume,arXiv:0811.0729,arXiv:0812.1440,arXiv:0901.3707,huang2010}.
Taking temperature $T\sim 350$ MeV, the relaxation time is around
$\t_\p\sim 0.27-1.35$ fm if using $\w/s=1/(4\p)-5/(4\p)$, where $s$
is the entropy density, and for free gluon gas it is
\begin{eqnarray}
\label{s} s&=&\n_g\frac{2\p^2T^3}{45},
\end{eqnarray}
with the degeneracy factor $\n_g=2(N_c^2-1)$.

Since the system has reflection symmetry with respect to the $yz$-plane,
 and there are no particle, momentum, or heat flow through the hard
 walls, the system obeys the following boundary conditions
\begin{eqnarray}
&v_z(t,0)=v_x(t,0)=v_x(t,\pm h)=0,&\non &\pt v_z(t,\pm h)/\pt x=0.&
\end{eqnarray}
As the initial state we choose uniform initial temperature of 355 MeV
 (corresponding to RHIC initial temperature), no flow in $x$-direction, and
 a simple sine-type longitudinal flow velocity profile
\begin{eqnarray}
&v_x(t_0,x)=0,&\non &v_z(t_0,x)=v_0 \sin{(\p x/2h)},&
\end{eqnarray}
where $v_0$ is the magnitude of the initial velocity at the two
 boundaries. In the following numerical calculation we consider two
 cases: $v_0 = 0.7$ and 0.9. In the expanding case, we use the initial time
 $\tau_0 = 1$ fm. Since the shear stress tensor is a dynamical variable in
 Israel-Stewart hydrodynamics, we need its initial value too. A natural
 choice is the Navier-Stokes value, but its exact evaluation is
 difficult. It contains the time derivative of the flow velocity, which
 is unknown before the hydrodynamic equation is solved. To avoid this
 problem, we initialize the shear stress, not to its exact Navier-Stokes,
 but to a ``static Navier-Stokes" value; that it, we ignore all the time
 derivatives in the Navier-Stokes definition of the shear stress tensor
 and calculate the value based on spatial derivatives only. In practice
 this means that some components of the tensor are slightly larger and
 some slightly smaller than their exact Navier-Stokes values.

\begin{figure}[!htb]
\begin{center}
\includegraphics[width=7cm]{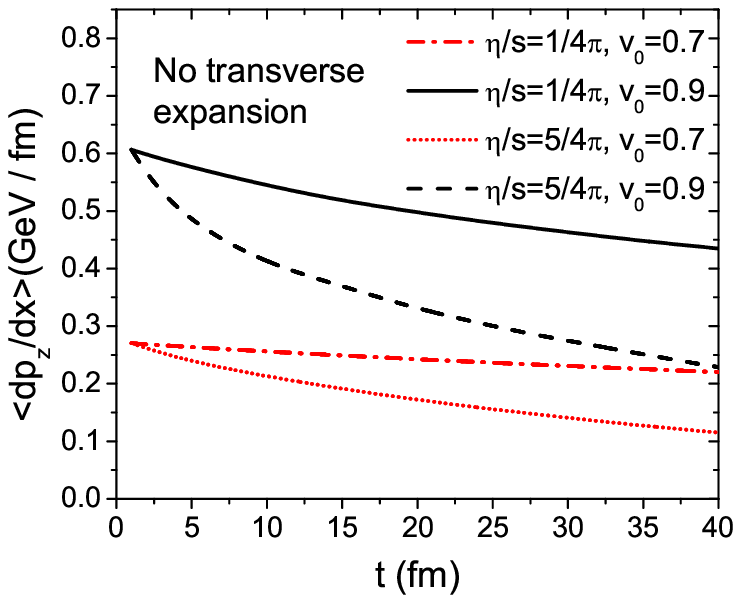}
\includegraphics[width=7cm]{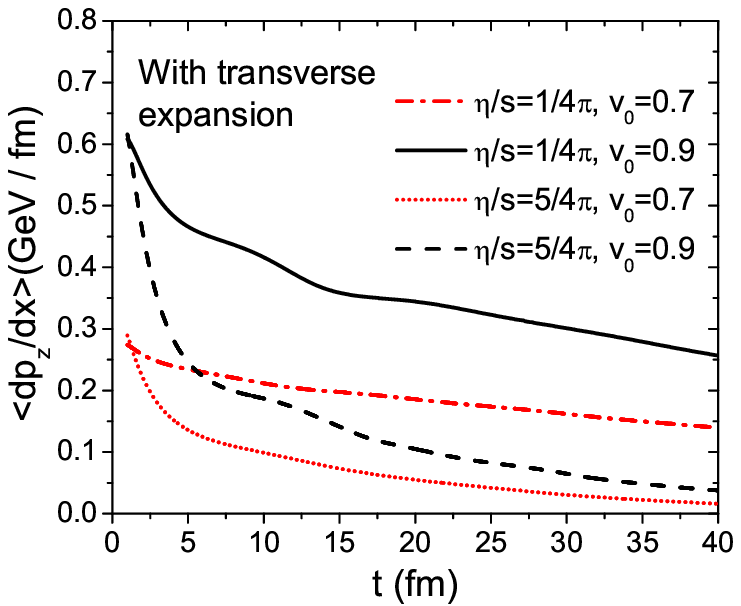}
\caption{(Color online) Evolution of the average gradient of the
longitudinal momentum per particle, $dp_z/dx$, at different shear
viscosities. Upper panel: the system has no transverse expansion.
Lower panel: the system has Bjorken expansion in the $\byh$ direction. }
\label{grad}
\end{center}
\end{figure}

In \fig{grad}, we depict the time evolution of the gradient of the longitudinal momentum per particle
averaged over $x\in[0,h]$,
\begin{eqnarray}
\Big\lan\frac{dp_z}{dx}\Big\ran\equiv\int_0^h dx
J^0(x)\frac{d}{dx}\frac{T^{0z}(x)}{J^0(x)}\Big/\int_0^h dx
J^0(x),\non
\label{eq:dpz}
\end{eqnarray}
where $J^0=\g\r$ is the proper particle number density. As expected, the
shear viscosity dissipates the average gradient of the longitudinal
momentum, especially for larger values of shear viscosity.
The transverse expansion accelerate this degradation, since
 strong transverse expansion means larger shear (shear tensor).
 

In the case of transverse expansion and large viscosity, there appear to be
a ``shoulder" in the time evolution of the longitudinal momentum
gradient $\lan dp_z/dx\ran$ as shown in the lower panel of \fig{grad}, 
where the gradient drops very fast  initially and then slows down for a while before
it decreases again. 
The temporary slow down is caused by the oscillatory behaviors of
 the induced transverse flow in $x$-direction, and the particle number
 density $J^0$, which is used as a weight in the calculation of the average
 longitudinal momentum gradient in Eq.~(\ref{eq:dpz}). The oscillations are an
 artifact of the fixed wall boundary conditions in our simple scenario.
When there is no transverse expansion, the degradation is slower and there is no shoulder because of the smaller 
shear (in shear tensor).


\begin{figure}[!htb]
\begin{center}
\includegraphics[width=8cm]{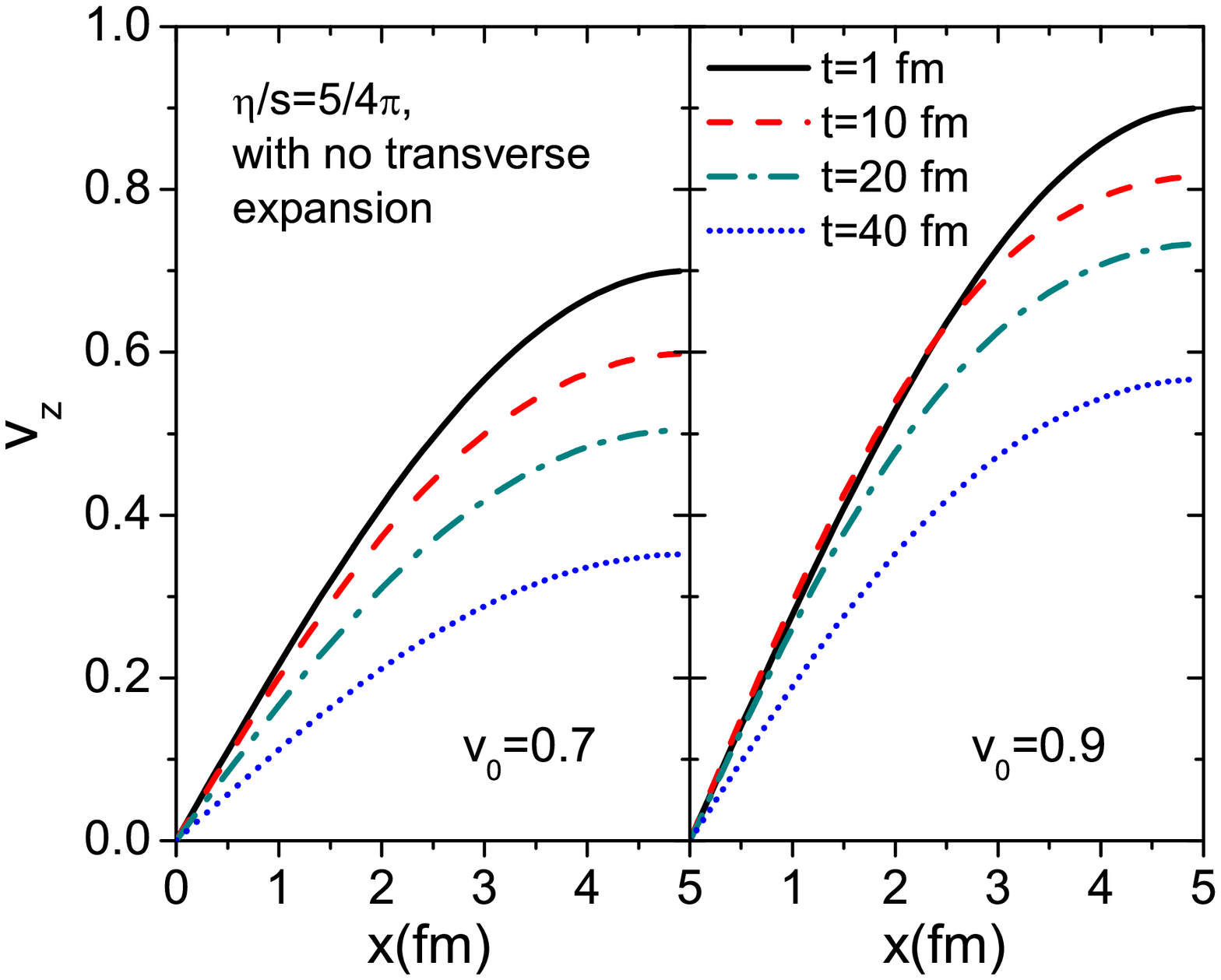}
\includegraphics[width=8cm]{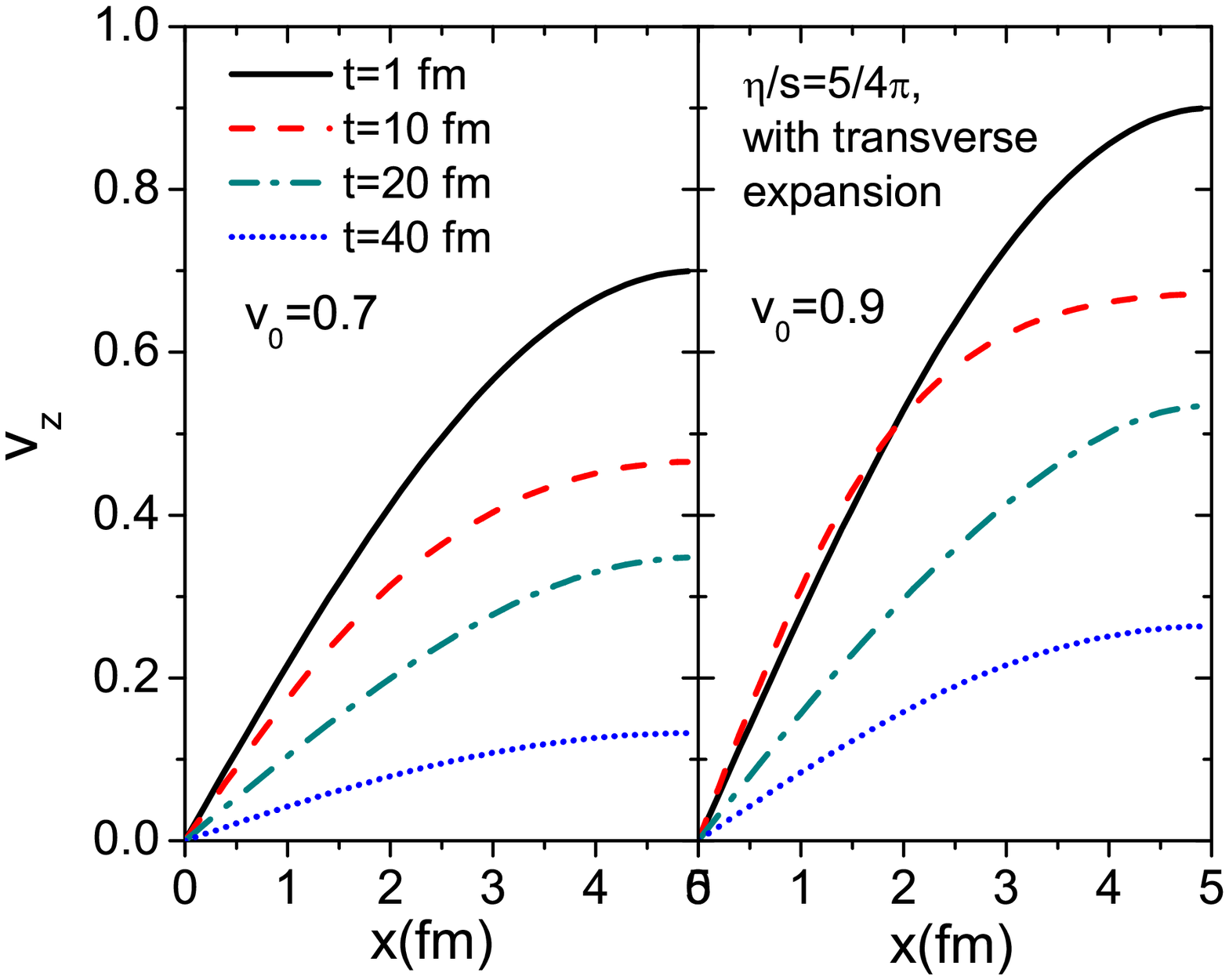}
\caption{(Color online) The profile of longitudinal velocity $v_z$
at different times with $\w/s=5/4\p$.} \label{vz_e5}
\end{center}
\end{figure}
\begin{figure}[!htb]
\begin{center}
\includegraphics[width=7cm]{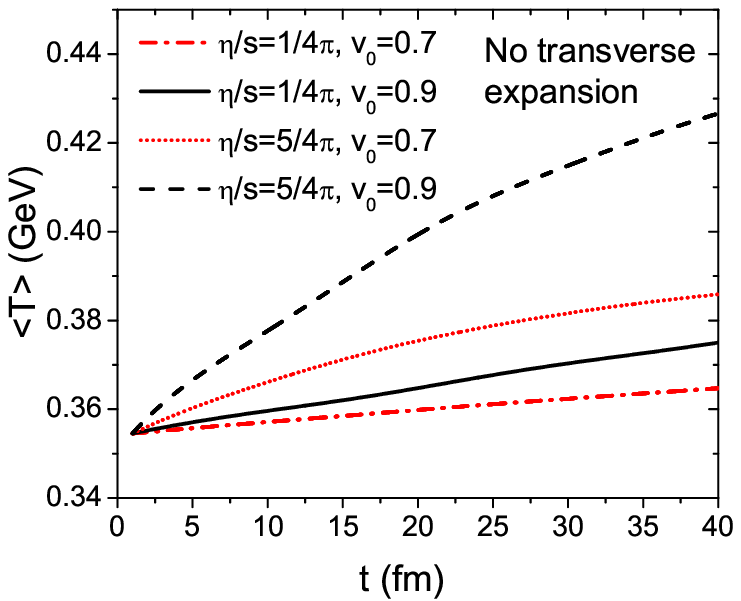}
\includegraphics[width=7cm]{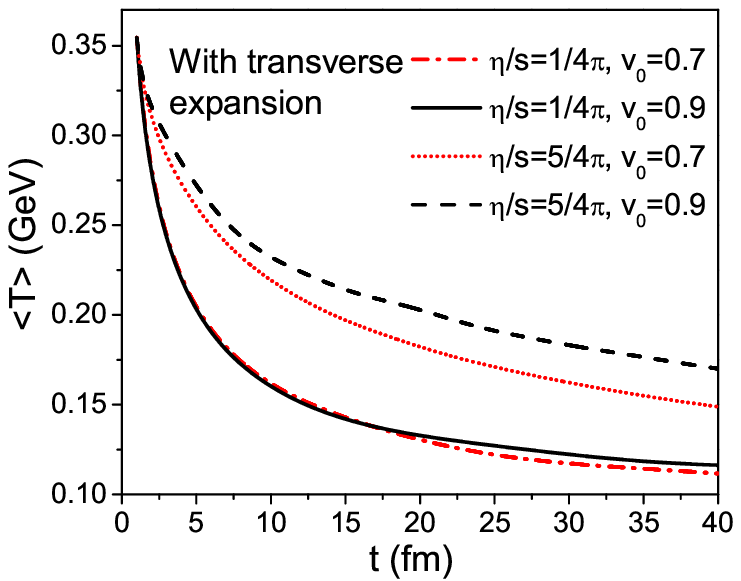}
\caption{(Color online) Evolution of the average temperature $\lan
T\ran$ with different shear viscosities and different initial
velocities. Upper panel: the system has no transverse expansion.
Lower panel: the system is Bjorken expanding in $\byh$ direction.}
\label{T}
\end{center}
\end{figure}

In \fig{vz_e5}, we show the profiles of velocity $v_z$ at different
time with viscosity $\w/s=5/4\p$ with (lower panel) and without (upper panel) transverse expansion.
One of the functions of the shear viscosity is to transform the kinetic
 energy of the fluid to internal energy, hence damping the fluid shear
 (as shown in Fig.~\ref{vz_e5}) and heating up the fluid. This can be explicitly
 seen in the upper panel of Fig.~\ref{T}, where the temperature evolution is
 shown for the nonexpanding system.
The transverse Bjorken expansion in our problem, however, will dilute the system
and cool the system down, overcoming the slight heating-up by the shear viscosity, as shown in the
lower panel of Fig.~\ref{T}. The transverse expansion will also accelerate the degradation of the
longitudinal velocity as shown in the lower panel of Fig.~\ref{vz_e5} as compared to
the upper panel for the case of no transverse expansion.


\section {Evolution of the Global Quark Polarization}\label{4}
With the model of time evolution of the longitudinal momentum gradient of the medium partons
we can now study the time evolution of the quark polarization when it is
progressively polarized due to multiple scattering.

According to Eq.~(\ref{pol}),  the change of polarization caused by one scattering is
\begin{eqnarray}
\D P&\equiv& P_f-P_i=-\frac{(1-P_i^2)\p\m p}{2E(E+m)-P_i\p\m p}.
\end{eqnarray}
For convenience we denote $P=P_i$. Then we get the following
evolution equation for the polarization,
\begin{eqnarray}
\frac{d P}{d t}&\equiv&\frac{\D
P}{\t_{q}}=-\frac{1}{\t_{q}}\frac{(1-P^2)\p\m p}{2E(E+m)-P\p\m p},
\end{eqnarray}
where $\t_{q}$ is the mean-free-path of quark which is related
to the transport cross section $\s_{tr}$ of the interacting partons
though $\t_q\simeq 1/(\r \s_{tr})$,  where $\r=\n_g\z(3)T^3/\p^2$ is
the density of medium gluons, assuming gluons are the dominant degrees of freedom in the medium.
The shear viscosity for a thermal ensemble of
gluons is roughly~\cite{Danielewicz:1984ww}
\begin{eqnarray}
\w\simeq\frac{1}{3}\r \lan p_{tr}\ran \frac{4}{9} \t_{q}\approx T\frac{4}{9}\r\t_q.
\end{eqnarray}
We have then the final rate equation for the time evolution of the quark polarization,
\begin{eqnarray}
\label{evoP} \frac{d P}{d
t}=-\frac{4T\r}{9s}\frac{s}{\w}\frac{(1-P^2)\p\m p}{2E(E+m)-P\p\m
p}.
\end{eqnarray}
\begin{figure}[!htb]
\begin{center}
\includegraphics[width=7cm]{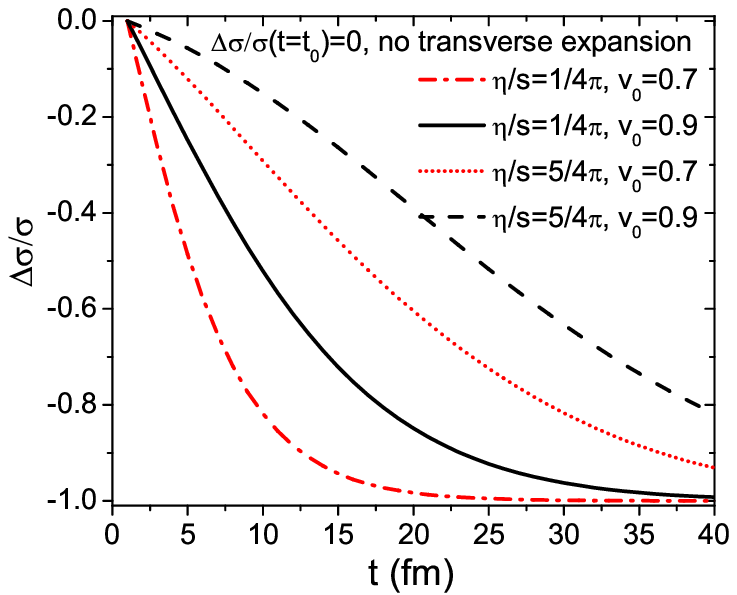}
\includegraphics[width=7cm]{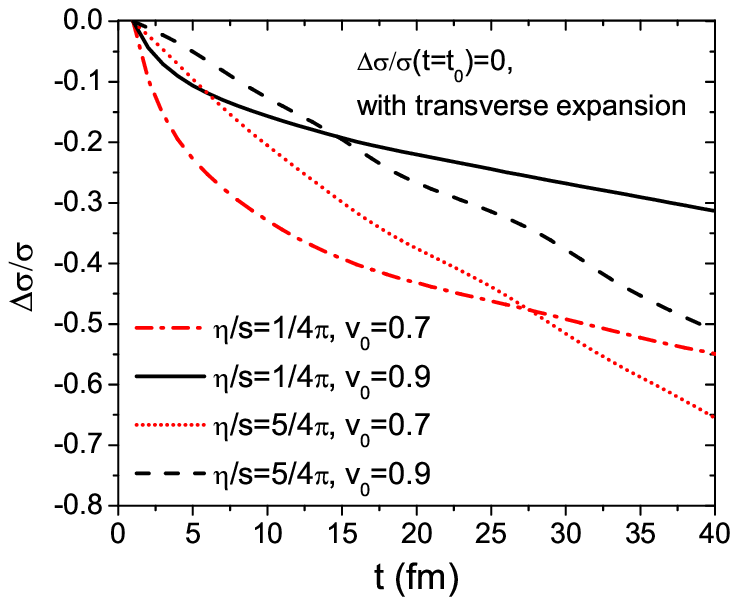}
\caption{(Color online) Evolution of the average polarization
$P=\D\s/\s$  with initial polarization
$P(t_0)=0$ with different values of viscosities
without (upper panel) and with (lower panel)
transverse expansion.} \label{po}
\end{center}
\end{figure}

From \eq{evoP}, the rate $dP/dt$ is inversely proportional to the
viscosity. This is evidently shown in the upper
panel of \fig{po}, in which the evolutions of the quark
polarizations are shown for the initial polarizations $P(0)=0$ and
for a system without transverse expansion.
With transverse expansion, the mean-free-path increases
more rapidly with time and therefore slows down the polarization
rate. The transverse expansion also accelerates the degradation of
the longitudinal momentum gradient, reducing the polarization
in each scattering. Both effects slow down the time evolution of
the polarization in an expanding system as shown by comparison between the
upper and lower panels of \fig{po}.

Because of the reheating by viscous interaction,
the initial cooling of the system due to transverse expansion is significantly slower
for a larger value of shear viscosity, as shown in Fig.~\ref{T}.  This speeds up the polarization
according to Eq.~(\ref{evoP}). However, a larger shear viscosity also slows down the polarization
because the polarization rate is inversely proportional to the shear viscosity. During the early stage
of evolution, the second effect dominates, leading to a slower polarization process with a larger
value of shear viscosity. At a later time, effect of reheating becomes more dominant and a larger shear viscosity
leads to a faster polarization process.

The polarization is also sensitive to the initial condition of the longitudinal flow shear. In our simple laminar flow
model, the initial longitudinal flow shear is proportional to the value of $v_0$. The nonlinear dependence of the polarization
rate on the relative momentum $p$ in Eq.~(\ref{evoP}) determines the nontrivial dependence of the polarization on the values
of $v_{0}$ as shown in Fig.~\ref{po}.

Note that the polarization rate we used are derived with the
approximation of small angle scattering which is only valid when the longitudinal momentum gradient is
large. For large shear viscosity $\eta/s$ and at late time, the longitudinal momentum
gradient can become too small. One can no longer use the rate equation derived here. However, one can 
assume that the polarization process will stop at this point when there is not
significant local orbital angular momentum.

\section{Summary}

In conclusion, we have calculated the polarization cross section for
quarks with initial polarization within the frame of perturbative QCD, which we use
to study the time evolution of the quark polarization via multiple scattering in a medium with nonvanishing
local orbital angular momentum. We considered
the simple case of laminar flow as governed by viscous hydrodynamics
with given shear viscosity $\eta/s$ and a simple illustrative initial condition.  Such a simple hydrodynamic model
provides the dynamic evolution of the longitudinal flow shear as the
polarization mechanism for quarks via parton scattering. Because the values of
the shear viscosity influence the degradation of the longitudinal flow shear
with time and the cooling of the system, it also determines the time
evolution of the quark polarization. Since the polarization rate is inversely proportional to the shear viscosity
and depends nonlinearly on the average longitudinal momentum shear, the final quark polarization
is found to be sensitive to the shear viscosity but has a nontrivial dependence. In this sense,
one can use the final state polarization as a
possible viscometer of the QGP.

For more realistic studies, one should employ a full scale 3+1D viscous hydrodynamics~\cite{Schenke:2010rr}
with initial conditions from Monte Carlo models such as HIJING~\cite{Wang:1991hta}. The initial parton
production from this kind of model has approximate Bjorken scaling which will give rise to very small
initial local longitudinal flow shear~\cite{arXiv:0710.2943} except at very large rapidity region.
Such small initial local longitudinal flow shear comes from the violation of the Bjorken scaling
which one can use as the initial condition. Furthermore, one should
also extend the current calculation of the quark polarization beyond the small angle
approximation.

\section*{Acknowledgments:}
We thank G.~Torrieri, D.~Rischke and Z.~Xu
for helpful discussions. This work is supported by the
Helmholtz International Center for FAIR within the framework of the
LOEWE (Landesoffensive zur Entwicklung
Wissenschaftlich-\"Okonomischer Exzellenz) program launched by the
State of Hesse,  by the ExtreMe Matter Institute (EMMI), and 
by BMBF under Contract No.\ 06FY9092
and by the director, Office of Energy Research,
Office of High Energy and Nuclear Physics, Divisions of Nuclear
Physics, of the U.S. Department of Energy under Contract No.
DE-AC02-05CH11231.
X.-N.~Wang thanks the hospitality of
the Institut f\"ur Theoretische Physik, Johann Wolfgang
Goethe-Universit\"at and support by 
EMMI in the framework of the Helmholtz Alliance Program of the
Helmholtz Association (HA216/EMMI) during the beginning of this
work.

\end{document}